\documentclass[fleqn,usenatbib]{mnras}

\usepackage{newtxtext,newtxmath}
\usepackage[T1]{fontenc}
\usepackage{ae,aecompl}
\usepackage{bm,url}

\usepackage{graphicx}	% Including figure files
\usepackage{amsmath}	% Advanced maths commands
\usepackage{amssymb}	% Extra maths symbols
\graphicspath{{./fig/}{./png/}}

\newcommand{\angles}[1]{\ensuremath{\left\langle #1 \right\rangle}}

\newcommand{\approptoinn}[2]{\mathrel{\vcenter{
	\offinterlineskip\halign{\hfil$##$\cr
	#1\propto\cr\noalign{\kern2pt}#1\sim\cr\noalign{\kern-2pt}}}}}

\usepackage{color}

\def\bl{Babcock--Leighton}
\def\mc{meridional circulation}

\newcommand{\Eq}[1]{Equation~(\ref{#1})}
\newcommand{\Eqs}[2]{equations~(\ref{#1}) and~(\ref{#2})}

\newcommand{\App}[1]{Appendix~\ref{#1}}
\newcommand{\Sec}[1]{\S\ref{#1}}

\newcommand{\Fig}[1]{Figure~\ref{#1}}
\newcommand{\Tab}[1]{Table~\ref{#1}}
\newcommand{\Figs}[2]{Figures~\ref{#1} and \ref{#2}}

\newcommand{\mps}{m~s$^{-1}$}

\newcommand{\cmss}{cm$^2$~s$^{-1}$}

\def\Rs{R_{s}}
\newcommand{\etaSCZ}{\eta_{\mathrm{SCZ}}}
\newcommand{\etaRZ}{\eta_{\mathrm{RZ}}}
\newcommand{\etasurf}{\eta_{\mathrm{surf}}}

\title[Stellar Dynamo with Solar and Anti-solar Differential Rotation]
{
Stellar Dynamos with Solar and Anti-solar Differential Rotations: Implications to Magnetic Cycles of Slowly Rotating Stars
}

\author[Karak, Tomar \& Vashishth]{
Bidya Binay Karak\thanks{E-mail: karak.phy@iitbhu.ac.in}, Aparna Tomar and Vindya Vashishth\\
$^{1}$Department of Physics, Indian Institute of Technology (Banaras Hindu University), Varanasi, India\\
}

\date{Accepted XXX. Received YYY; in original form ZZZ}

\pubyear{2019}

\begin{document}
\label{firstpage}
\pagerange{\pageref{firstpage}--\pageref{lastpage}}
\maketitle

\begin{abstract}
Simulations of magnetohydrodynamics convection in slowly rotating stars predict anti-solar differential rotation (DR) in
which the equator rotates slower than poles.
This anti-solar DR in the usual $\alpha \Omega$ dynamo model does not produce
polarity reversal. Thus, the features of large-scale magnetic fields in slowly rotating stars 
are expected to be different than stars having solar-like DR. 
In this study, we perform mean-field kinematic
dynamo modelling of different stars at different rotation periods. We consider anti-solar DR for the stars
having rotation period larger than 30~days and solar-like DR otherwise. We show that with particular
$\alpha$ profiles, the dynamo model produces magnetic cycles with polarity reversals even with the anti-solar DR
provided, the DR is quenched when the toroidal field grows considerably high and there is a sufficiently strong
$\alpha$ for the generation of toroidal field. Due to the anti-solar DR, the model produces an abrupt increase of magnetic field exactly when the DR profile is changed from solar-like to anti-solar. 
This enhancement of magnetic field is in good agreement with the stellar observational data as well as some global convection simulations. In the solar-like DR branch, with the decreasing rotation period, we find the magnetic field strength increases while the cycle period shortens. Both of these trends are in general agreement with observations. Our study provides additional support for the possible existence of anti-solar DR in slowly rotating stars and the presence of unusually enhanced magnetic fields and possibly cycles which are prone to production of superflare.
\end{abstract}

\begin{keywords}
Sun: activity, dynamo, magnetic fields --- stars: solar-type, rotation, activity.
\end{keywords}

%%%%%%%%%%%%%%%%%%%%%%%%%%%%%%%%%%%%%%%%%%%%%%%%%%

%%%%%%%%%%%%%%%%% BODY OF PAPER %%%%%%%%%%%%%%%%%%

\section{Introduction}
\label{sec:int}
Many sun-like low-main sequence stars show magnetic cycles which are usually studied by measuring the chromospheric
emissions in Ca~II H \& K line cores and the coronal X-ray \citep{Baliu95,Pal81}. These measurements
show that the periods of magnetic cycles vary in different stars and there is a weak trend 
of the period becoming shorter with the increase of rotation rates \citep{Noyes84b,Suarez16,BoroSaikia18}.
Some recent observations, however, do not find this 
%weak trend; see Fig.\ 21 of \citet{Suarez16}.
trend in the so-called active branch; see Fig.\ 21 of \citet{Suarez16}.
On the other hand, the stellar magnetic activity or the field strength increases with the increase of rotation rate
in the small rotation range and then it tends to saturate in the rapid rotation limit.
This behaviour is often represented with respect to the Coriolis number (or the inverse of Rossby number) 
which is a ratio of the convective turnover time to the rotation period \citep{Noyes84a,wright11,WD16}. However,
\citet{RSP14} show that rotation period alone better represents the data.
In the slowly rotating stars with rotation rates below about the solar value, 
an interesting behaviour has been recognised.
\citet{Giamp06,Giamp17} have found an increase of magnetic activity with a decrease of $\tau/P_{\rm rot}$
(i.e., increase of rotation period). 
Another exciting aspect of stellar activity is that the magnetic cycles of 
Sun-like stars show the Waldmeier effect (the stronger cycles rise
faster than the weaker ones) \citep{garg19}, which is popularly known for the Sun \citep{wald, KC11}.
%Recently, \citet{BG18} highlighted this feature further and they have argued that this increased magnetic activity
%is due to the change in the differential rotation (hereafter DR) profile from solar-like to the anti-solar in the slowly rotating stars. 

\begin{figure}
\centering
\includegraphics[width=.95\columnwidth]{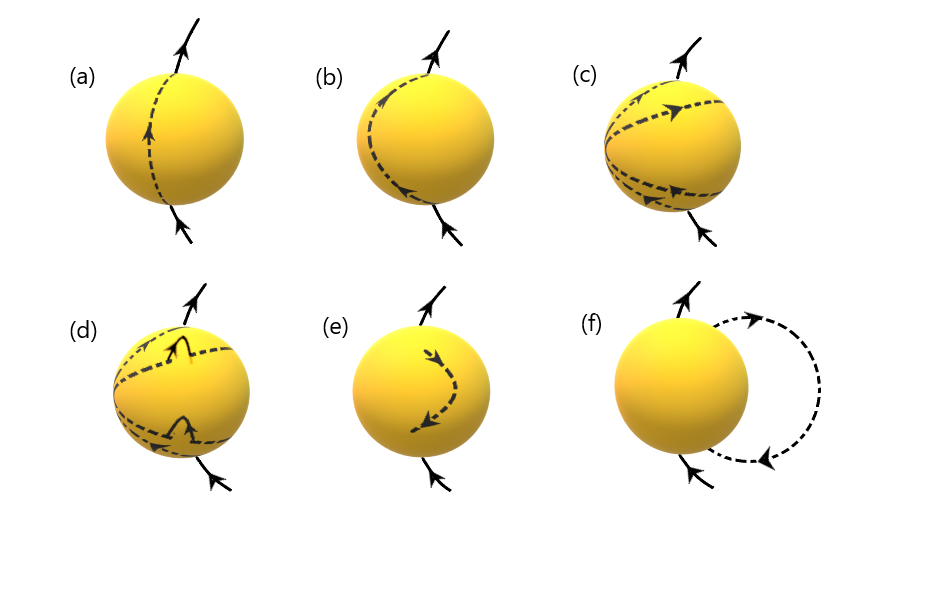}
\caption{
Pictorial representation of the \bl\ dynamo model with anti-solar DR.
(a) Initial poloidal field line. (b) and (c) This field is stretched by the anti-solar DR 
(equator rotates slower than poles) to produce a toroidal field.
(d) The toroidal field rises to the surface and form tilted BMRs with loop structures.
(e) Opposite polarity of sunspots connects near the equator and create a big poloidal loop (dashed line).
(f) The large-scale poloidal field developed (dashed line) is in the same orientation as that of the original one.}
\label{fig:antisol}
\end{figure}

Starting from the pioneering work of \citet{Gi77,Gi83}, numerous global magnetohydrodynamics (MHD) convection simulations
in spherical geometry have
been performed to study the magnetic fields and flows in stars. 
These simulations in some parameter ranges
produce large-scale magnetic fields and even cycles. In general, it has been observed that in the rapidly rotating stars,
magnetic cycles and polarity reversals are preferred, while in the slowly rotating ones, simulations rarely produce
reversals \citep{BB17,W18}.
Most of the simulations near the solar rotation rate
(or Rossby number around one) find a transition
of differential rotation (hereafter DR) from solar-like to the so-called anti-solar profile, in which
the equator rotates slower than high latitudes \citep{Gi77,Gue13,gastin,Kap14,FF14,Kar15,FM15,KMB18}.
In most of the  cases, when simulations produce anti-solar DR, they do not produce magnetic field reversals \citep{W18}.
It is not difficult to understand that the anti-solar DR in $\alpha \Omega$ dynamo model, does not allow polarity reversal. The anti-solar DR generates a toroidal field in such a way that
the poloidal field produced through the $\alpha$ effect (positive in the northern hemisphere) from this toroidal field
is in the same direction as that of
the old poloidal field. In \Fig{fig:antisol}, we show how a \bl\ type $\alpha \Omega$ dynamo with anti-solar
DR produces poloidal field in the same direction as that of the original field and thereby does not offer
reversal of the poloidal field at the end of a dynamo cycle.

Based on the available data of the observed polar field, DR and other surface features,
it is expected that the solar dynamo is primarily of $\alpha \Omega$ type \citep{CS15}. In recent years, it
has been realized that it is the \bl\ process which acts like an $\alpha$ effect to generate the poloidal field
through the decay and dispersal of tilted bipolar magnetic regions (BMRs) \citep{Das10,KO11,Muno13,Priy14}. Although in the Sun, DR
is the dominating source of the toroidal field, a weak toroidal field might be generated through the
$\alpha$ effect. With the increase of stellar rotation rate, $\alpha$ increases \citep{KR80}, 
while the DR ($\Omega$) does not have a strong dependency with the rotation \citep{KR99}. 
Thus essentially, the stellar dynamo is of $\alpha^2 \Omega$ type
and this type of dynamo can result in polarity reversal depending on the profiles of $\alpha$ and $\Omega$.
Therefore, in the simulations with anti-solar DR,
the magnetic field reversal is subtle as the toroidal field can be produced through
the $\alpha$ effect in addition to $\Omega$. Incidentally, \citet{Kar15} found little reversal of magnetic field
although it does not occur globally in all latitudes; see their Fig.\ 9 and 10, Runs A--BC.
They also found some irregular cycles in anti-solar DR regime. Recently, \citet{viv18}
found noticeable polarity reversal in anti-solar DR regime, however, \citet{W18} found almost no signature
of polarity reversals in this regime. In summary, there is no consensus in the polarity reversal
of the large-scale magnetic field in the simulations of the slowly-rotating stars producing anti-solar
DR. In this study, we shall explore this and understand how the magnetic polarity reversal and cycles can be possible in this regime.

The most interesting feature is the change of magnetic field strength in the slowly rotating stars.
As referenced above, observations of slowly rotating stars show an increase of magnetic activity
with the increase of rotation period which is possibly caused by the transition of DR to anti-solar from the solar-like profile as proposed by \citet{BG18}.
Indeed, \citet{Kar15} found an increase of the magnetic field in the anti-solar DR regime. 
Simulations of \citet{W18} in different parameter regimes seem to show a little increase of the magnetic field, 
in contrast, \citet{viv18} did not find significant increase.
This remains a curiosity in the community whether the magnetic field indeed increases in the anti-solar DR regime
and if so, then whether this is actually caused by the anti-solar DR. 
The answers to these are not obvious
because the strength of the dynamo, as determined by the $\alpha$ and shear, do not necessarily increase in the anti-solar DR regime (slowly rotating stars). In fact, we expect the strength of $\alpha$ to decrease with the decrease of rotation rate.
Well, in simulations it has been observed that the shear is much strong in the anti-solar DR regime (see Table~1 of \citet{Kar15} and \citet{viv18}). 
But this cannot be the reason for the increased magnetic activity in slowly rotating stars because
otherwise both \citet{viv18} and \citet{W18} would also find an abrupt increase of magnetic field.
Therefore, we shall understand the behaviour of the stellar dynamo with anti-solar DR which is found in the slowly rotating stars.
We shall explore whether the increase of magnetic field as seen in observation is possible in the dynamo model with anti-solar
DR and what causes this increase.

In our study, without going through the complexity of the global MHD convection simulations
and saving the computational resources, we shall develop a simple kinematic dynamo model 
by specifying $\alpha$ and $\Omega$ for the sources of magnetic fields 
and make some clean simulations to explore the physics
behind the magnetic cycles and polarity reversal in stars. We shall explore how the magnetic field strength varies 
when the DR profile changes from solar to anti-solar.
We shall also present how other features of magnetic cycles, namely cycle period and the ratio of poloidal to toroidal fields change with the stellar rotation.

\section{Model}
\label{sec:mod}
In our study, we develop a kinematic mean-field dynamo model by considering only the diagonal 
terms of the $\alpha$ coefficients and an isotropic turbulent diffusion $\eta$. 
We further assume magnetic field to be axisymmetric, 
thus writing magnetic field $\bm{B} = \bm{\nabla}\times [A (r,\theta,t) \bm{\hat{\phi}}] + B (r,\theta,t) \bm{\hat{\phi}}$, where $A$ is vector potential for poloidal field $\bm{B_p} \equiv (B_r \bm{\hat{r}},B_\theta \bm{\hat{\theta}})$, $B$ is toroidal field, and $\theta$ is co-latitude. 
Subsequently, the equations for $A$ and $B$ can be derived as

\begin{equation}
\frac{\partial A}{\partial t} + \frac{1}{s}({\bm v_p}\cdot \bm \nabla)(s A)
= \eta \left( \nabla^2 - \frac{1}{s^2} \right) A + \alpha_{\phi\phi}B
\label{eq:pol}
\end{equation}

\begin{eqnarray}
\frac{\partial B}{\partial t}
+ \frac{1}{r} \left[ \frac{\partial}{\partial r}
(r v_r B) + \frac{\partial}{\partial \theta}(v_{\theta} B) \right]
= \eta \left( \nabla^2 - \frac{1}{s^2} \right)B \nonumber \\
+ s(\bm B_p\cdot{\bm \nabla})\Omega + S^{\rm Tor}_\alpha(r,\theta)+
\frac{1}{r}\frac{d\eta}{dr}\frac{\partial{(rB)}}{\partial{r}}
\label{eq:tor}
\end{eqnarray}\\
where $s = r \sin \theta, {\bm v_p} \equiv (v_r\bm{\hat{r}},v_\theta \bm{\hat{\theta}}))$, and
\begin{eqnarray}
S^{\rm Tor}_\alpha(r,\theta) = -\alpha_{\theta\theta}\left(\frac{2}{r}\frac{\partial A}{\partial r} + \frac{\partial^2 A}{\partial r^2} \right)\nonumber \\
+\frac{\alpha_{rr}}{r^2}\left(\frac{A}{\sin^2\theta}-\cot\theta\frac{\partial A}{\partial \theta}
-\frac{\partial^2 A}{\partial \theta^2} \right)\nonumber \\ 
-\frac{1}{r}\frac{\partial \alpha_{rr}}{\partial \theta} \left(\frac{A \cot\theta}{r} + \frac{1}{r}\frac{\partial A}{\partial \theta}\right)
-\frac{\partial \alpha_{\theta\theta}}{\partial r} \left(\frac{A}{r}+\frac{\partial A}{\partial r}\right)
\label{eq:Sal}
\end{eqnarray}
with $\alpha_{rr}, \alpha_{\theta\theta}$ and $\alpha_{\phi\phi} $ being the diagonal components of $\alpha$ tensor. 
In our model, we shall ignore the off-diagonal components
(which are responsible for magnetic pumping) because of the limited knowledge of their profiles in the solar and stellar convection zones (CZs) and to keep our model and the
interpretation of the results tractable.

\begin{figure}
\centering
\includegraphics[width=1.0\columnwidth]{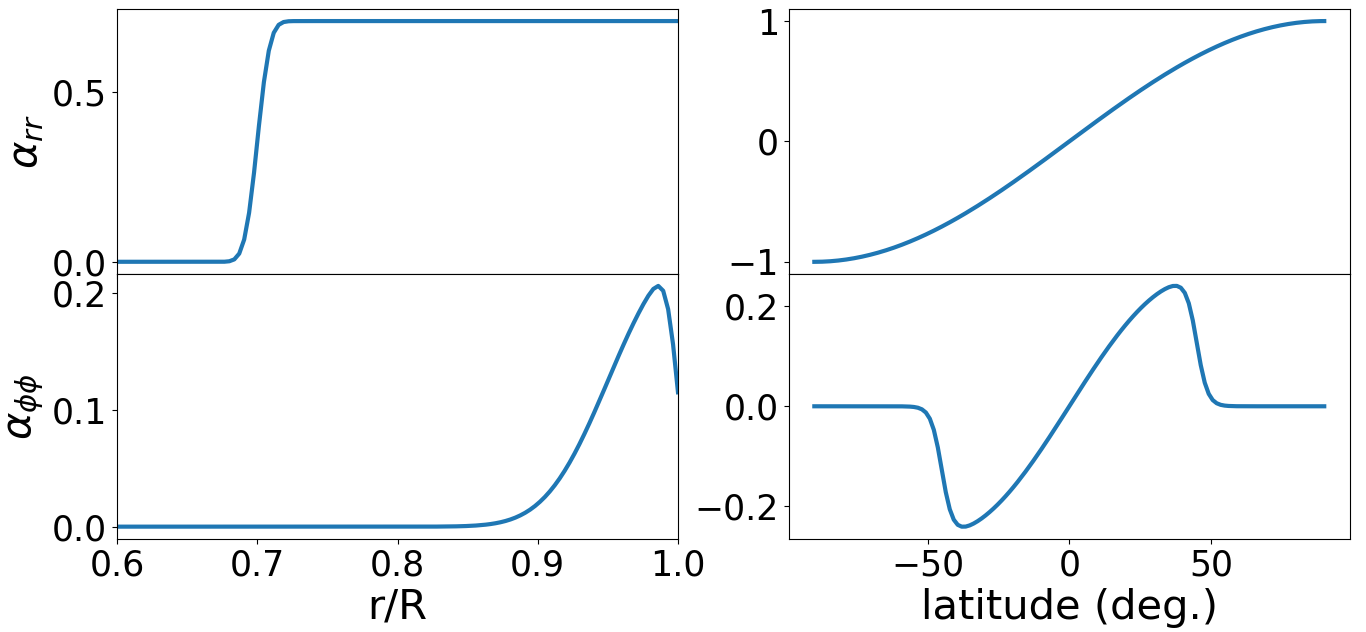}
\caption{
Left: Radial variations of $\alpha_{rr}$ (\Eq{eq:alr}) and $\alpha_{\phi\phi}$ (\Eq{eq:alp}) at $45^\circ$ latitude for $\alpha_0=1$~\mps.
Right: Latitudinal variations of the same at $r=0.95\Rs$.
}
\label{fig:alpha}
\end{figure}

Numerical simulations of magneto-convection in local Cartesian domain \citep[e.g.,][]{KB09,Kar14b}
and global solar/stellar CZs \citep[e.g.,][]{sim13,sim16,War18} provide 
some guidance, although the simulation carried out are still far from the real Sun 
(in terms of its fundamental parameters). These simulations show that the $\alpha$ 
components are highly inhomogeneous across radius and latitudes. They change sign 
at the equator and have no resemblance with each other. We also note that there 
is some amount of uncertainty in the measurement of $\alpha$ coefficients, and so far, 
we do not have a well-tested method for their measurement. Thus keeping general features 
of the $\alpha$ coefficients as obtained through theory and simulations, we begin with the 
following simple profiles for $\alpha_{rr}$ and $\alpha_{\theta\theta}$.
\begin{equation}
\alpha_{rr}=\alpha_{0}\frac{1}{2}\left[1+\mathrm{erf}\left(\frac{r-0.7\Rs}{0.01\Rs}\right)\right]\cos\theta,
\label{eq:alr}
\end{equation}
\begin{equation}
\alpha_{\theta\theta}=\alpha_{rr},
\label{eq:alt}
\end{equation}
where $\Rs$ is the solar radius; 
see \Fig{fig:alpha} top panels for the variations of this profile.

For $\alpha_{\phi\phi}$, we use a different profile 
(\Fig{fig:alpha} bottom panels) and it is given by
\begin{eqnarray}
\alpha_{\phi\phi}=\alpha_{0}\frac{1}{4}\left[1+\mathrm{erf}\left(\frac{r-0.95\Rs}{0.05\Rs}\right)\right]\left[1-\mathrm{erf}\left(\frac{r-\Rs}{0.01\Rs}\right)\right] \nonumber \\
 \times f_{\rm{s}} \cos\theta\sin\theta, ~~~~~~~~~~~~
\label{eq:alp}
\end{eqnarray}
where $f_{\rm{s}}$ is a function that takes care of suppressing $\alpha_{\phi\phi}$ above $\pm45^\circ$ latitudes and thus 
 $f_{\rm{s}} = 1 / [1 + \exp\{30(\pi/4-\theta)\}]$ for $\theta<\pi/2$ and 
$1/[1 + \exp\{30(\theta - 3\pi/4)\}]$ for otherwise. We take this factor $f_{\rm{s}}$ in $\alpha_{\phi\phi}$ to restrict the strong toroidal field and thus the band of formation of sunspots in the low latitudes, which is a common practice in the flux transport dynamo models \citep{KRS01, Dik04, GD07, HY10, KC16}.
Further, we choose $\alpha_{\phi\phi}$ to operate only near the surface so that the source for the poloidal field is slightly segregated from the source for the toroidal field. 
This, in turn, helps in producing longer cycle period of 11~years.
%Following the general strategy of Babcock-Leighton dynamo models, we have $\alpha_{\phi\phi}$ non zero only in low latitudes which helps to keep strong toroidal field in the low latitudes and is in agreement with the fact that sunspots are formed only in the low latitudes.
For $\alpha_{rr}$ and $\alpha_{\theta\theta}$ we keep the profiles simple so that they both are non-zero in the whole CZ and have a $\cos\theta$ dependence. 
Later in \Sec{sec:robust}, we shall also use different $\alpha$ profiles and 
explore the robustness of our results.

To limit the growth of magnetic field in our kinematic dynamo model, 
we consider following simple nonlinear quenching
\begin{equation}
%\alpha_{ij} = \frac{\alpha_{ij}}{1+\left(\frac{B}{B_{0}}\right)^2}
\alpha_{ii} = \frac{\alpha_{ii}}{1+\left(\frac{B}{B_{0}}\right)^2}
\end{equation}
where $i = r, \theta, \phi$ and $B_{0}$ is the saturation field strength which is fixed at $4\times 10^4$~G 
in all the simulations. Because of this nonlinear saturation,
we always present the magnetic field from our model with respect to $B_{0}$.

\begin{figure}
\centering
\includegraphics[width=.85\columnwidth]{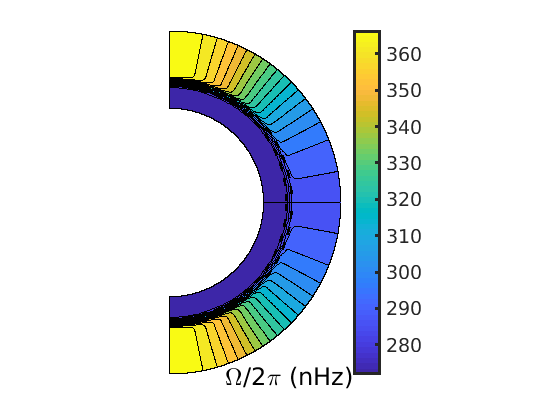}
\caption{
Angular frequency $\Omega/2\pi$ for the anti-solar DR; see \Eq{eq:DR}.
}
\label{fig:antiDR}
\end{figure}

For DR, we take following profile which reasonably fits the helioseismology 
data
\begin{equation}
\Omega(r,\theta) = \Omega_\mathrm{RZ} + \frac{1}{2}\left[1 + \mathrm{erf} \left(\frac{r - 0.7\Rs}{0.025\Rs}\right) \right] \left(\Omega_\mathrm{CZ} - \Omega_\mathrm{RZ}\right),
\label{eq:DR}
\end{equation}
where $\Omega_\mathrm{RZ}/2\pi = 432.8$~nHz, and
$\Omega_\mathrm{CZ}/2\pi = 460.7 - 62.69\cos^2\theta -  67.13\cos^4\theta$~nHz.
When we make the DR anti-solar, we take
$\Omega_\mathrm{CZ}/2\pi = 460.7 + 62.69\cos^2\theta +  67.13\cos^4\theta$~nHz.
Note that this will allow $\Omega$ to increase with latitudes and that is the
anti-solar DR which is shown in \Fig{fig:antiDR}.

The turbulent magnetic diffusivity $\eta$ has the following form:
\begin{equation}
\eta(r) = \etaRZ + \frac{\etaSCZ}{2}\left[1 + \mathrm{erf} \left(\frac{r - 0.7\Rs}
{0.02\Rs}\right) \right]+\frac{\etasurf}{2}\left[1 + \mathrm{erf} \left(\frac{r - 0.9\Rs}
{0.02\Rs}\right) \right],
\label{eq:eta}
\end{equation}
where 
$\etaRZ = 5\times10^{8}$~\cmss, 
$\etaSCZ = 5\times10^{10}$~\cmss, and
$\etasurf =  2\times10^{12}$~\cmss.
%Meridional circulation ($v_r$ and $v_\theta$) and turbulent diffusivity ($\eta$) profiles are 
Meridional circulation ($v_r$ and $v_\theta$) profile is  
the same as given in \citet{HY10}; also see \citet{Dik04,KKC14, KP13}. For the sake of completeness, 
we write these profiles and boundary conditions in \App{sec:appen}.

\section{Results}
\label{sec:res}

We start all simulations by specifying some initial magnetic field, and we analyse the results
only after running the code for several diffusion times so that the magnetic field
reaches a statistically stationary state. We first present the result of an $\alpha^2$ model
by setting $\Omega=0$. As the results for such a dynamo model for the Sun with parameters given above
has not been presented before, we shall first show the result.

\subsection{Solar dynamo without differential rotation}
\label{sec:sun}
\begin{figure}
\centering
\includegraphics[width=1.1\columnwidth]{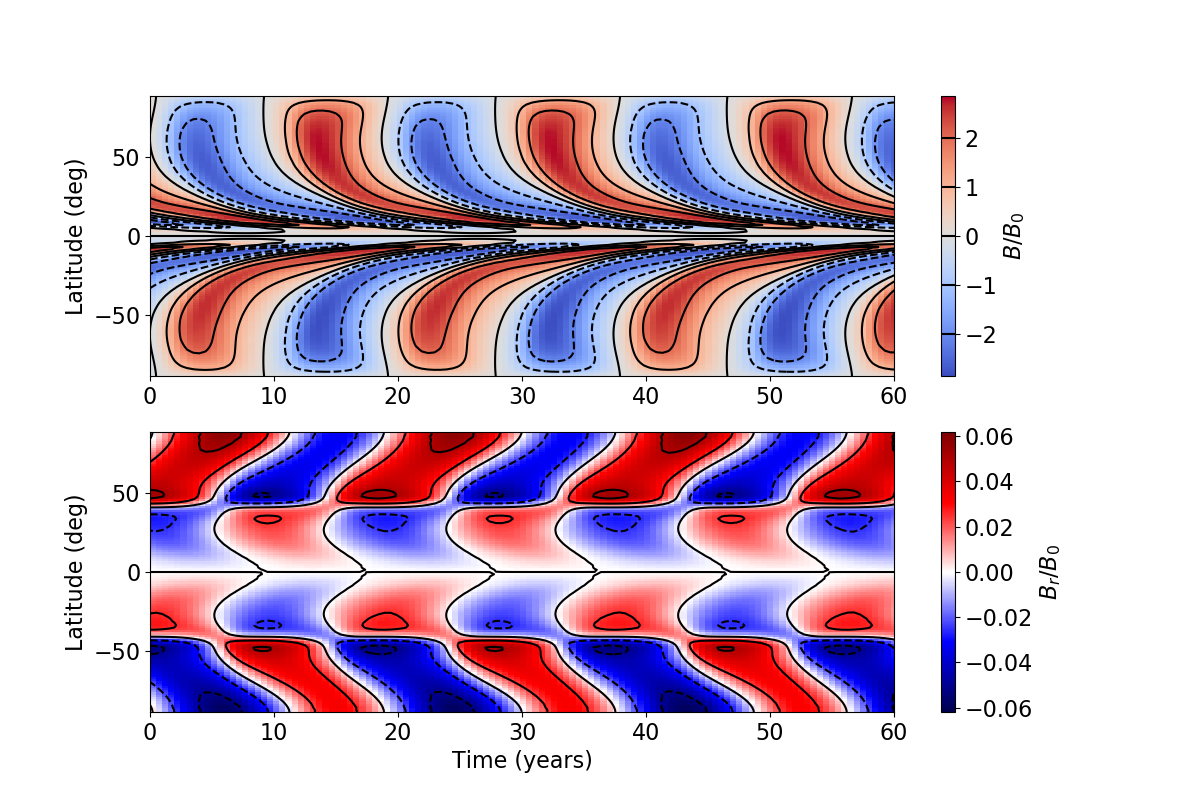}
\caption{
Result of $\alpha^2$ dynamo model with meridional circulation.
Time-latitude diagrams of (a) toroidal magnetic field $B$ at $r=0.7\Rs$
and surface radial field $B_r$. All fields are measured with respect to $B_0$.
}
\label{fig:alphasq}
\end{figure}

The critical $\alpha_0$ for the $\alpha^2$ dynamo model without DR
for the Sun is about $10$~\mps. However, the prominent dynamo cycles with polarity reversals
are seen only when $\alpha_0$ is approximately above 18~\mps.
Butterfly diagram for $\alpha_0 = 20$~\mps\ is shown in \Fig{fig:alphasq}. 
We notice that even this model without DR
produces some basic features of the solar cycle, namely, (i) the polarity
reversal, (ii) 11-year periodicity, (iii) equatorward migration of toroidal field at the base of the CZ, and (iv) the poleward
migration of surface poloidal field.  Feature (i) is due to the selected 
inhomogeneous profile of $\alpha$
taken in our model. 
The other properties are largely determined by the inclusion of meridional flow which is poleward near the surface and equatorward near the base of the CZ and a relatively weak diffusivity in the bulk of the CZ.

The oscillatory magnetic field in $\alpha^2$ dynamo in spherical geometry is not new. 
It was realized that the $\alpha^2$ dynamo with certain $\alpha$ profiles 
\citep{BS87,SG03,ER07}, and/or boundary conditions \citep{Mi10,Ja17} produces 
an oscillatory solution. We also note that the mean-field model
constructed using $\alpha$ profiles obtained from the global MHD simulation also finds
oscillatory solution \citep{sim13}. 
%A similar study was also performed by \cite{viv19}, although they did not perform a mean-field model.

\subsection{Stellar dynamo with solar and anti-solar DR}
\label{sec:stars}
Now we perform dynamo simulations of different stars with rotation period 1, 5, 10, 17, 25.38 (solar value), 30, 32, 40 and 50 days. 
As discussed in the introduction that the slowly rotating stars are expected to have anti-solar DR, 
which is, at least, confirmed in the global MHD simulations. This transition from solar to anti-solar DR is possibly happening below the solar rotation with Rossby number not too far from unity.
Therefore, we assume that all stars with rotation periods up to 30 days are having solar-like DR, 
while stars of periods longer than 30 days have anti-solar DR.
Further, the amount of internal DR is expected to change with the rotation rate. 
Hence, we choose it to scale in the following way.
\begin{equation}
\Omega (r,\theta) = \left(\frac{T_s}{T}\right)^n \Omega_s (r,\theta),
\label{eq:Omegascale}
\end{equation}
where $T_s = 25.38$~days (solar rotation period), $T$ is the rotation period of star, 
and $\Omega_s (r,\theta)$ is the internal angular frequency of the Sun 
as given in \Eq{eq:DR}. Some numerical simulations suggest $n$ is about 0.3 \citep{bal07,Brown08}, 
while the recent work of \citet{viv18} in a more wider range finds $n\approx -0.08$ 
for the rotation rates up to 5 times solar value and 
$n = -0.96$ for rotation rate of 5--31 times the solar rotation.
Surface observations find somewhat similar results, 
for example, \citet{barnes05}, \citet{RG15}, \citet{DSB96}, and \citet{Le16} 
respectively give $n=0.15, 0.29, 0.7,$ and $-0.36$. 
In our study, we perform two sets of simulations by considering two values,
namely $n=0.7$, i.e., the DR increases
with the increase of rotation rate with an exponent of $0.7$ and $n=0.0$, i.e.
no change in DR. These two sets of simulations are labelled
as A1--A50 and B1--B50; see \Tab{table1}.

\begin{table}
\caption{Summary of simulations. Here, $T$ is the rotation period of star in days, $P_{\rm cyc}$ is the mean magnetic cycle period, SL and AS stand for solar and anti-solar DR, respectively.
}
\begin{center}
\begin{tabular}{llccccc}
\hline
Details  & Run & $T$ (d) & DR & $\frac{B_{\rm tor}}{B_0}$ & $\frac{B_{\rm pol}}{B_0}$ & $P_{\rm cyc}$ (yr)\\

\hline
Set A:     & A1&   ~1     &SL&  6.346  &   0.329 &  ....\\
$\Omega$ and $\alpha$    
           & A5&   ~5     &SL&  1.636  &   0.066 &  ~....\\
change     & A10&   10     &SL&  0.663  &   0.051 &  ~3.30\\
following  & A17&   17     &SL&  0.361  &   0.035 &  ~2.00\\
Eqs.
           & A25&   25.38  &SL&  0.141  &   0.028 &  ~7.41\\
\ref{eq:Omegascale}--\ref{eq:alphascale}.
           & A30&   30     &SL&  0.128  &   0.024 &  ~7.60\\
           & A32&   32     &AS&  0.582  &   0.081 &  22.95\\
           & A40&   40     &AS&  0.623  &   0.066 &  20.95\\
           & A50&   50     &AS&  0.575  &   0.053 &  18.52\\
\hline
Set B:     & B1&   ~1     &SL&  5.885  &  0.340 &  ....\\
Same as    & B5&   ~5     &SL&  1.702  &  0.071 &  ~2.10\\
Set A but  & B10&   10     &SL&  0.355  &  0.053 &  ~3.40\\
no change  & B17&   17     &SL&  0.326  &  0.037 &  ~7.92\\
in DR.     & B25&   25.38  &SL&  0.140  &  0.023 &  ~7.38\\
           & B30&   30     &SL&  0.213  &  0.023 &  ~7.20\\
           & B32&   32     &AS&  0.680  &  0.081 &  24.31\\
           & B40&   40     &AS&  0.748  &  0.066 &  21.13\\
           & B50&   50     &AS&  0.708  &  0.054 &  19.41\\
\hline
Set C:     & C1&   ~1     &SL&  3.628 &  0.146 & ....\\
Same as    & C5&   ~5     &SL&  1.534 &  0.057 & ~2.54\\
Set A but  & C10&   10     &SL&  0.370 &  0.048 & ~6.53\\
meridional & C17&   17     &SL&  0.399 &  0.036 & ~2.82\\
flow       & C25&   25.38  &SL&  0.141 &  0.028 & ~7.41\\
changes    & C30&   30     &SL&  0.062 &  0.010 & ~7.51\\
following  & C32&   32     &AS&  0.603 &  0.082 & 19.96\\
Eqs. \ref{eq:mcscale}
           & C40&   40     &AS&  0.701 &  0.070 & 15.52\\
           & C50&   50     &AS&  0.251 &  0.024 & 12.19\\
\hline
Set D:     & D1&   ~1     &SL&  0.762 &  0.0319 & 10.50\\
Same as    & D5&   ~5     &SL&  0.287 &  0.0051 & 9.206\\
Set A but  & D10&   10     &SL&  0.261 &  0.0054 & 12.49\\
$\alpha_{rr}$, $\alpha_{\theta\theta}$
           & D17&   17     &SL&  0.186 &  0.0041 & 12.33\\
are taken  & D25&   25.38  &SL&  0.133 &  0.0030 & 12.33\\
from Eqs.
           & D30&   30     &SL&  0.110 &  0.0025 & 12.24\\
\ref{eq:alrD}--\ref{eq:altD}
           & D32&   32     &AS&  0.119 &  0.0032 & 14.09\\
           & D40&   40     &AS&  0.115 &  0.0033 & 14.23\\
           & D50&   50     &AS&  0.110 &  0.0035 & 14.22\\
\hline
Set E:
     & E1&   ~1      &SL&  4.578 &  1.956  & ....\\
Same as    & E5&   ~5      &SL&  1.999 &  0.8728 &~0.54\\
Set A but  & E10&   10     &SL&  1.266 &  0.5821 &~1.86\\
all $\alpha$ are
           & E17&   17     &SL&  1.113 &  0.4765 & ~2.23\\
taken from & E25&   25.38  &SL&  0.573 &  0.293 & ~2.92\\
Eq.\ \ref{eq:alrE}  
           & E30&   30     &SL&  0.445 &  0.250 & ~2.95\\
and high 
           & E32&   32     &AS&  0.450 &  0.355 & ~7.93\\
diffusivity.
           & E40&   40     &AS&  0.300 &  0.313 & ~4.90\\
           & E50&   50     &AS&  0.233 &  0.272 & ~4.83\\
\hline
\end{tabular}
\end{center}
\label{table1}
\end{table}

\begin{figure}
\centering
\includegraphics[width=1.0\columnwidth]{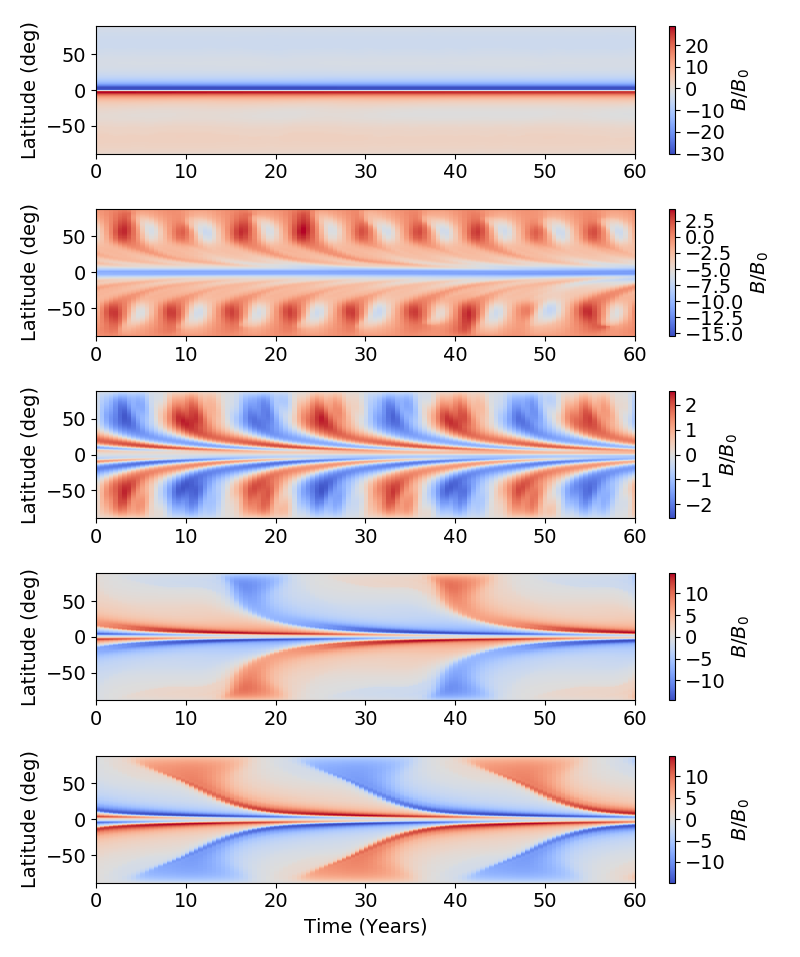}
\caption{
Butterfly diagram of the toroidal field averaged over a thickness of $0.04\Rs$ 
centered at $0.71\Rs$
from stars of rotation periods 5, 10, 25.38 (Sun), 32, and 50 days 
(top to bottom); Runs~A5, A10, A25, A32, and A50.
}
\label{fig:allbfly}
\end{figure}

\begin{figure}
\centering
\includegraphics[width=1.0\columnwidth]{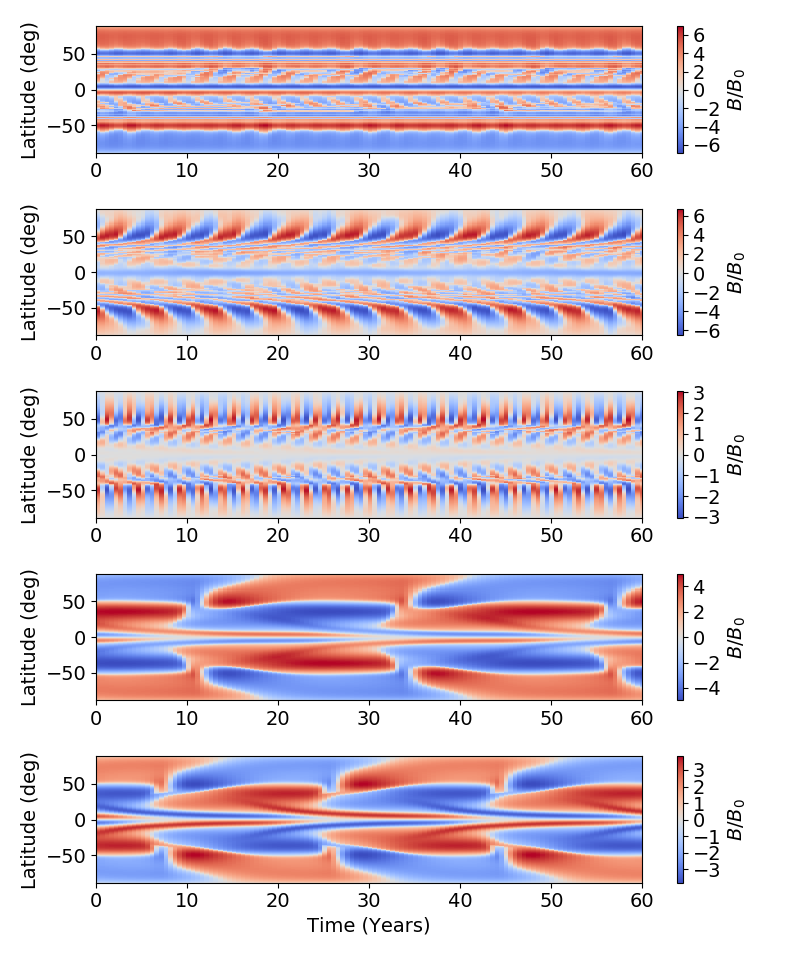}
\caption{
Same as \Fig{fig:allbfly} but obtained from the toroidal field at $0.85\Rs$.
}
\label{fig:allbflym}
\end{figure}

\begin{figure}
\centering
\includegraphics[width=1.\columnwidth]{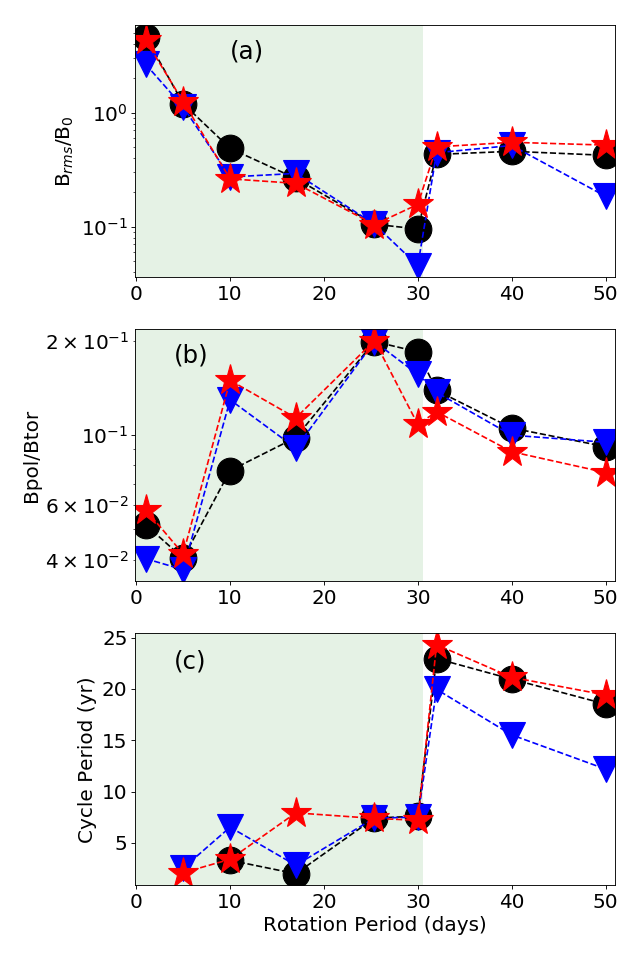}
\caption{
Top--bottom: Dependences of the magnetic field, the ratio of toroidal to poloidal fields,
and cycle periods with rotation periods of the stars.
Circular and asterisk points are obtained from simulations in which DR
decreases with rotation period through \Eq{eq:Omegascale} (Runs~A1--A50) and
DR does not change (Runs~B1--B50), respectively.
Triangular points come from the same simulations as that of circular points but the meridional circulation is increased with rotation period through \Eq{eq:mcscale} (Runs~C1--C50).
The shaded area represents the solar-like DR regime, while the white space is anti-solar.
}
\label{fig:BvsT}
\end{figure}

The amplitude of $\alpha$ is also expected to increase with the increase of rotation
rate \citep{KR80}. 
As we do not know how exactly $\alpha$ scales with the rotation rate,
we choose it to increase linearly. 
Thus, in Equations~(\ref{eq:alr}--\ref{eq:alp}), 
we take
\begin{equation}
\alpha_0 = \frac{T_s}{T} \alpha_{0,s},
\label{eq:alphascale}
\end{equation}
where $\alpha_{0,s}$ is the value of $\alpha_0$ for the solar case, in which
we take $\alpha_{0,s}=80$~\mps.

The \mc\ is another important ingredient in the model which is expected to vary 
with the rotation rate. 
%Our knowledge of \mc\ even for the sun is very limited. Theoretical calculations and 
Our knowledge of \mc\ even for the sun is very limited. Mean-field models and 
global convection simulations
all suggest that its profile as well as amplitude changes with the rotation 
rate \citep{KO11b,FM15,Kar15,KMB18}.
Simulations of \citet{Brown08} showed that the kinetic energy of \mc\
approximately scales as $\Omega^{-0.9}$. Therefore in one set of simulations
(Runs~C1--C50 in \Tab{table1}), we shall change \mc\ along with other parameters in the following way.
\begin{equation}
v_0 = \left(\frac{T_s}{T}\right)^{-0.45} v_{0,s},
\label{eq:mcscale}
\end{equation}
where $v_{0,s}$ is the amplitude of \mc\ of sun, for which we have taken 10~\mps.
We do not make any other changes in the model. 

Interestingly, all the stars, including ones with anti-solar DR (rotation periods 32--50~days), 
show prominent dynamo cycles and polarity reversals except for the very rapidly rotating
star of rotation period 1~day. The butterfly diagrams of stars with rotation periods 5, 10, 25.38,
32, and 50 days (from Runs~A1--A50) are shown in \Figs{fig:allbfly}{fig:allbflym}.
We observe that the 5~days rotating star (Run A5) does not show magnetic cycle 
at the base of the CZ,
but it does some cycles in low latitudes of mid-CZ. The magnetic fields are largely
anti-symmetric across the equator in all the cases. 
The equatorward migration is largely due to the return meridional flow.

To quantify how the magnetic field strength and cycle period change
with different stars, we compute the root-mean-square ($rms$) field strength
$B_{rms} = \sqrt{\angles{(B_r^2+B_\theta^2+B_\phi^2)}}$,
the ratio of poloidal to toroidal field Bpol/Btor 
$= \sqrt{\angles{(B_r^2+B_\theta^2)}} /\sqrt{\angles{B_\phi^2}}$ 
(where angular brackets denote the average over the whole domain), and the mean 
period of the polarity reversals of the toroidal field.
These quantities are listed in \Tab{table1} and the variations with rotation period are shown in \Fig{fig:BvsT}.

In the rapidly rotating regime, the magnetic field strength increases 
with the decrease of rotation period,
which is in agreement with observations \citep{Petit08}.
This behaviour in fact is not very surprising because the magnetic field sources ($\alpha$ and $\Omega$) 
are increased with the rotation rates through \Eqs{eq:alphascale}{eq:Omegascale}. 
However, what is surprising in \Fig{fig:BvsT}(a) is that the magnetic field abruptly increased 
just above the rotation period of 30 days. 
That is the point where the DR pattern is changed from solar-like to anti-solar.
Thus the anti-solar DR causes a sudden increase of magnetic field in the slowly rotating stars,
which is in good agreement with the stellar observational data \citep{Giamp06,Giamp17,BG18} 
and also with global MHD convection simulations \citep{Kar15}.
Further, we note that even the simulations (Run~B1--B50; red asterisks in \Fig{fig:BvsT}a)
in which $\Omega$ does not change with the rotation rate (i.e., $n=0$ in \Eq{eq:Omegascale})
also produces a similar variation of magnetic field. 
As the shear is not decreased in this case,
the field in the anti-solar branch is little
stronger in comparison to Set~A (black points).
The Set~C, in which meridional flow is increased with the rotation period (\Eq{eq:mcscale}),
also show a similar variation of magnetic field (blue triangles in \Fig{fig:BvsT}a), which is unusual in the $\alpha\Omega$ 
type flux transport dynamo models \citep{Kar10}.

\begin{figure}
\centering
\includegraphics[width=1.2\columnwidth]{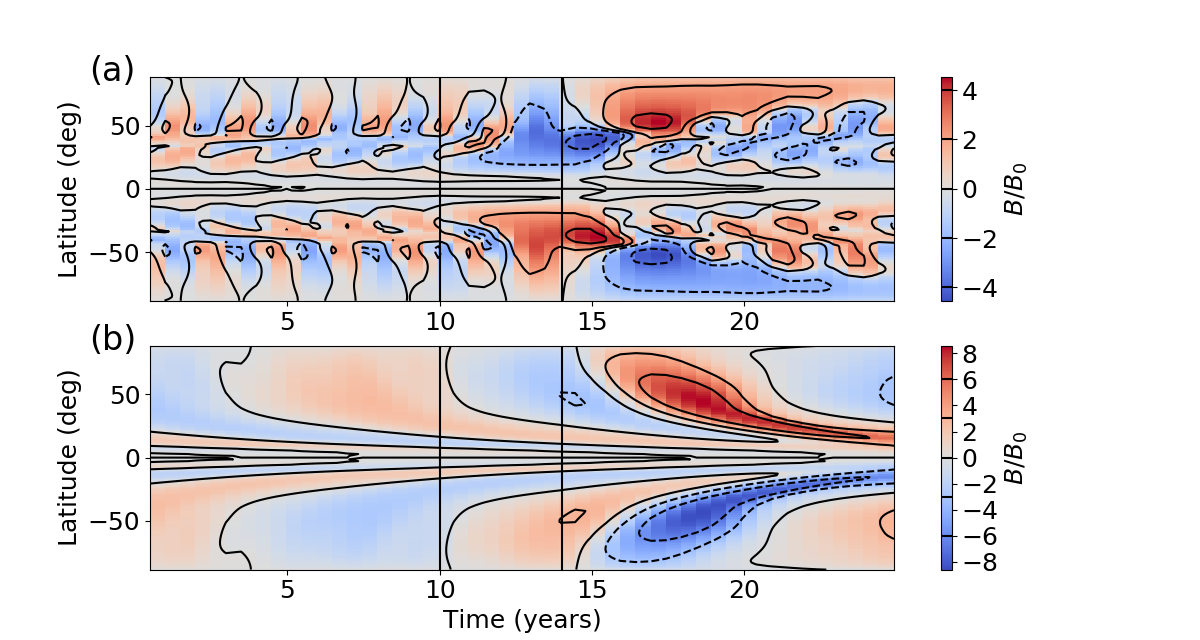}
\caption{Simulation result of a star with rotation period 30~days in which 
the DR is made anti-solar during 10--14~years (identified by two vertical lines).
Butterfly diagram of (a) the toroidal field at $0.85\Rs$
and (b) the same but averaged over a thickness of $0.04\Rs$ centered at $0.71\Rs$.
}
\label{fig:idea}
\end{figure}

%\subsection{Understanding the stellar magnetic cycles}
%\label{sec:understand}
%We shall first discuss how the anti-solar DR produces an enhanced magnetic field.
%To do so, we perform the following experiment. We consider the simulation of a star with rotation 
To understand the enhancement of the magnetic field in the anti-solar DR regime, we
perform the following experiment. We consider the simulation of the 30-day rotation period 
having solar-like DR (Run~A30). %While the model is producing steady dynamo solution,
We first make sure that this model is producing a steady dynamo solution.
Then, when the toroidal field reverses, that is when the toroidal field is at a minimum, we stop the code.
The vertical line at $t=10$~years in \Fig{fig:idea} shows this time.
Then we make the DR anti-solar and run the code for
four years (up to the second vertical line in \Fig{fig:idea}).
Finally, we revert the DR to the solar-like profile and continue the run for another 10~years.
We see a significant increase of magnetic field in \Fig{fig:idea}(a) after the DR is made anti-solar.
This increase of field must be caused by the change in DR from solar to anti-solar profile.
We know that during the reversal, the anti-solar DR produces toroidal field of the same sign
as that of the previous cycle. As seen in \Fig{fig:idea}(a), when the toroidal field in the northern hemisphere is mostly negative during the phase of anti-solar DR ($t=$10--14~years),
the dynamo needs a positive field to reverse it. However, the anti-solar DR gives
negative polarity field, and thus, it tries to enhance the old field (negative in the northern hemisphere). 
So essentially what is happening is the following:\\
%\begin{equation}
%\rm {Pol}(+) \overrightarrow{\rm{Solar-like DR}} \rm {Tor}(-) \rightarrow  
%\end{equation}
Pol($+$) ~$\overrightarrow{(Solar~\Omega)}$~ Tor($-$) ~$\overrightarrow{\alpha}$~
Pol($-$) ~$\overrightarrow{(Anti{-}sol~\Omega)}$~ Tor($-$).\\
Note in the last phase, the toroidal field is produced in the same polarity as that of the 
previous cycle and thus the field is enhanced; also see \Fig{fig:antisol}.
We may mention that instead of anti-solar DR if the $\alpha$ is reversed, then also the same effect will be seen. In fact, following this idea, recently \citet{Kar18} explained the sudden 
increase of the poloidal field and appearance of double peaks in the sunspot cycles 
using a momentarily reversed $\alpha$ due to wrong BMR tilt.

Later, when the DR is changed to solar-like, this increasing effect of anti-solar DR is
stopped and the model succeed to reverse the field (at around $t = 16$~years in \Fig{fig:idea}).
We note that the increase of magnetic field is not immediately seen in the averaged field 
at the base of the CZ in \Fig{fig:idea}(b) because of the cancellation and finite transport time, 
but it is clearly seen in the next cycle. 

\begin{figure}
\centering
\includegraphics[width=1.\columnwidth]{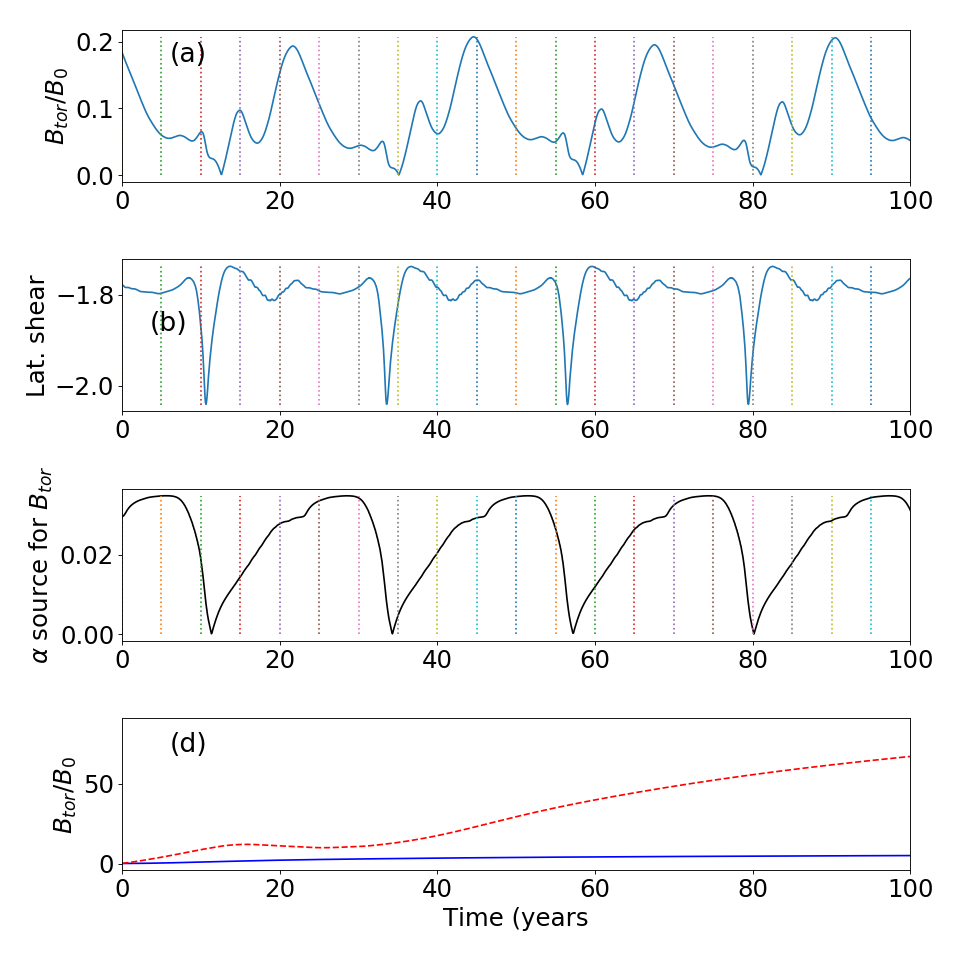}
\caption{Results from Run~A32 computed over the whole CZ in the northern hemisphere: 
Temporal variations of (a) the toroidal field,
(b) the mean effective latitudinal shear $\frac{1}{r}\frac{\partial\Omega}{\partial\theta}/\left(1+(B/B_0)^2\right)$
(in unit of $10^{-9}$ nHz~deg$^{-1}$), (c) the mean source term due to $\alpha$
effect for the toroidal field ($S^{\rm Tor}_\alpha$ in \Eq{eq:tor} in unit of $10^{-6}$s$^{-1}$), and
(d) the same as (a), but obtained from simulations in which
there is no quenching in the shear (red dashed line) 
and $\alpha_0 =  20$~\mps\ (solid blue), instead of 80~\mps\ 
which is used in all Runs~A1--A50.
}
\label{fig:shear}
\end{figure}

Above discussion shows that the anti-solar DR amplifies the old toroidal field
by supplying toroidal field of the same polarity. Thus we do not expect polarity reversal
if the anti-solar DR acts for all the time and if there is no other source for the toroidal
field generation. In our model, we certainly have $\alpha$ effect for the generation of the
toroidal field in addition to $\Omega$. However, this may not be sufficient to reverse the field.
We realized that when the toroidal field has reached a very high value, 
the DR is needed to be suppressed otherwise the $\alpha$ effect is not able to reverse the toroidal field and the cyclic dynamo is not possible. When the toroidal field is sufficiently high due to anti-solar DR, it is expected that this strong field acts back on the DR through Lorentz forces and tries to suppress the shear. 
The suppression mechanism could be complicated as the strong magnetic field can give Lorentz
forces on the large-scale flows as well as suppress 
the convective angular momentum transport \citep[see e.g,][]{Kar15,Kap16}.
The bottom line is that the shear must be decreased once the toroidal field reaches 
a sufficiently high value. This, in fact, is seen in the simulations of \citet{Kar15} 
that the shear parameters are quenched when the magnetic field is strong; see their Fig.\ 13(e-f).
We also note that even in the Sun, there is an indication of the reduction of shear during the 
solar maximum; see Fig.\ 6--7 of \citet{ABC08} and also \citet{BSG16}.
Motivated by these, we have introduced a nonlinear magnetic field dependent quenching
$f_q = 1/\left(1+(B/B_0)^2\right)$ in the shear such that $\bm\nabla\Omega$ in \Eq{eq:tor} is replaced by
$f_q\bm\nabla\Omega$.
Through this nonlinear quenching, the toroidal field tries to reduce the shear when it exceeds $B_0$. 
In \Fig{fig:shear}(b), we observe that the shear is strong during cycle minimum
and thus, it produces a strong toroidal field. Then the strong field quenches the shear and the toroidal field cannot grow rapidly. Interestingly, the toroidal field generation due to     
$\alpha$ effect $S^{\rm Tor}_\alpha$ (see \Eq{eq:Sal}) is not negligible (\Fig{fig:shear}(c))
and it is this process which eventually dominates and reverses the toroidal field slowly.
Essentially, it is the nonlinear competition between the toroidal field generation through 
shear and $\alpha$, which makes the polarity reversal possible even with the anti-solar DR.

To further support above conclusion, we show that if we do not include the quenching (i.e., $f_q = 1$) in the shear, then the dynamo fails to reverse the magnetic field, rather the toroidal field increases in time; see red/dashed line in \Fig{fig:shear}(d). On the other hand, if we keep
the quenching in shear but reduce the strength of $\alpha$ by $75\%$, then also, dynamo fails to
reverse the cycle and the magnetic field remains steady; see blue/solid line  in \Fig{fig:shear}(d).
In conclusion, to obtain the polarity reversal with anti-solar DR, there must be a sufficiently strong $\alpha$ and the shear must be reduced when the toroidal field becomes very strong.

Returning to \Fig{fig:BvsT}(b), we observe that for all the stars, the mean poloidal field
is about one order of magnitude smaller
than that of the toroidal field.
The ratio Bpol/Btor has a non-monotonous behaviour. It is maximum at the solar rotation
and decreases on both sides. This implies that the toroidal field dominates over the poloidal field
both for rapidly and slowly rotating stars.
With the increase of rotation rate, the toroidal field must increase faster than the poloidal component
because there are two sources of toroidal field ($\alpha$ and shear) and both increase with the rotation
rate. In the anti-solar branch, Bpol/Btor decreases because toroidal field generation is stronger
with the anti-solar DR.
The increase of poloidal contribution with the increase of cycle period in the solar-like DR
branch
is in agreement with the observational findings \citep{Petit08}.
These observational results also
show that the magnetic energy of the toroidal field
dominates below the rotation period of 12 days, while in our simulations the toroidal energy
is always greater than the poloidal field.
We should not forget that observational results are based on the field what was detected and it could be that most of the toroidal field
in observations was not be detected. 
Therefore, instead of comparing the actual value of magnetic energy, we should see how the ratio of poloidal to toroidal field changes with the rotation rate and this is in somewhat agreement with observation (compare Figure 6 of \citet{Petit08} with the left branch of our \Fig{fig:BvsT}(b)).
Unfortunately, \citet{Petit08} do not have data in the slowly rotating branch above the solar rotation
period and thus we cannot compare this behaviour with observations.

Finally, the cycle period again shows a non-monotonous behaviour. It decreases
with the increase of rotation rate, which is in general agreement with observations \citep{Noyes84b,Suarez16,BoroSaikia18},
although the observational trend is messy.
The slow decrease of rotation period in the anti-solar DR branch is not apparent in the observed
data; see, for example, Fig.\ 9 of \citet{BoroSaikia18}.
%We note that in the traditional flux transport dynamo model, 
In our model, the decrease of cycle period with the increase of rotation rate is because the dynamo becomes stronger with the rotation rate which makes the conversion between poloidal and toroidal faster and reduces the cycle period.
In global MHD simulations of stellar dynamos, \citet{Gue18} also find a decrease of period with the increase of rotation rate, 
while \citet{W18} find this trend only at slow rotation and then an increasing trend in the rapid rotation. 
In latter simulations, the increase of period with rotation rate is due to the decrease of shear.
In our study, when the DR is anti-solar, it tries to produce the same polarity field as that of
the previous cycle and thus $\alpha$ takes longer time to reverse the field---causing 
the cycle longer.
This effect decreases when the rotation period is too long because the strength of 
the anti-solar DR decreases and thus the polarity reversal becomes faster. 
This causes the cycle period to shorten with the increase of rotation period in the anti-solar branch.

%In all the above simulations, the \mc\ is kept constant which may not be justified. 
%Our knowledge of \mc\ even for the sun is very limited. Theoretical calculations and 
%global convection simulations
%suggest that its profile as well as amplitude changes with the rotation rate \citep{KO11b,FM15,Kar15}.
%Following the simulations of \citet{Brown08}, in which the kinetic energy of \mc\
%approximately scales with $\Omega$ as $\Omega^{-0.9}$, we consider
%\begin{equation}
%v_0 = (T_s/T)^{-0.45} v_{0,s},
%\end{equation}
%where $v_{0,s}$ is the amplitude of \mc\ of sun, for which we have taken 10~\mps.
%In runs C1--C50, we have decreased \mc\ with the rotation rate in this form 
%and the triangular points show the results of these simulations.

As the meridional flow transports the magnetic fields from source regions, the cycle duration
tends to be longer with the decrease of meridional flow \citep{DC99,Kar10}. This was usually seen in previous flux transport dynamo models of stellar cycles \citep{Nan04,JBB10,KKC14}, but not in the turbulent pumping-dominated regime \citep{KC16,Hazra19}.
Therefore, in Set~C in which \mc\ is decreased with the rotation rate following \Eq{eq:mcscale},
we expected an increase of cycle period. However, we see
a reverse trend (see blue points in \Fig{fig:BvsT}(c)), 
because the dynamo becomes stronger and this tries to make the cycle shorter.

\section{Robustness of our results}
\label{sec:robust}
We have above seen that the results are only little sensitive to the change in $\Omega$ and 
\mc. Notably, the enhancement of the magnetic field at the transition point from solar
to anti-solar DR is prominent. To explore the robustness of these results further, we 
consider different profiles for $\alpha$. In principle, we can consider countless different profiles for
$\alpha$ as the actual profiles are not known even for the Sun. Different profiles will make the 
dynamo solutions different but not necessarily the variations of magnetic field with rotation period. 
To demonstrate this, we present the results for two more sets of 
simulations, namely, Set D (Runs~D1--D50) and set E (Runs~E1--E50); see \Tab{table1}. 

In Set D, we change $\alpha_{rr}$ and $\alpha_{\theta\theta}$
to as follows:
\begin{equation}
\alpha_{rr}=\alpha_{0}\sin\left[2\pi\left(\frac{r-0.7\Rs}{0.3\Rs}\right)\right]\cos\theta\sin^2\theta,
\label{eq:alrD}
\end{equation}
\begin{equation}
\alpha_{\theta\theta}=\alpha_{0}\sin\left[2\pi\left(\frac{r-0.7\Rs}{0.3\Rs}\right)\right]\cos\theta,
\label{eq:altD}
\end{equation}
both for $r\ge 0.7\Rs$, while $\alpha_{rr}=\alpha_{\theta\theta}=0$ for $r<0.7\Rs$ and $\alpha_{0}=100$~\mps; 
see \Fig{fig:alphaD} for these profiles.
No other parameters are changed in this Set.
%We note that this type of profile are suggested in 

On the other hand, in Set E, we change all three $\alpha$ profiles to followings.
\begin{equation}
%\alpha_{rr}=\alpha_{\theta\theta}=\alpha_{\phi\phi}=\alpha_{0}\frac{1}{2}\left[1+\tanh\left(\frac{r-0.7\Rs}{0.015\Rs}\right)\right]\cos\theta,
\alpha_{rr}=\alpha_{\theta\theta}=\alpha_{\phi\phi}=\alpha_{0}\frac{1}{2}\left[1+\mathrm{erf}\left(\frac{r-0.7\Rs}{0.01\Rs}\right)\right]\cos\theta.
\label{eq:alrE}
\end{equation}
%With these new $\alpha$, we, however, do not get polarity reversal.  Interestingly, 
%in addition to this change in $\alpha$, when we change the diffusivity to as
%follows, then we obtain polarity reversal in the anit-solar DR branch.
%\begin{equation}
%\eta = \eta_\mathrm{RZ}+0.45\eta_\mathrm{CZ} \left[1+\tanh\left(\frac{r-0.7\Rs}{0.015\Rs}\right)\right],
%\label{eq:etaE}
%\end{equation}
%where $\eta_\mathrm{CZ} = 10^{12}$~\cmss\ and $\eta_\mathrm{RZ}=0.1\eta_\mathrm{CZ}$.
%Note that this type of $\alpha$ and $\eta$ profiles are obtained from \citet{KKT06}.
%in addition to this change in $\alpha$, when we change the diffusivity to as
%follows, then we obtain polarity reversal in the anit-solar DR branch.
%\begin{equation}
%\eta = \eta_\mathrm{RZ}+0.45\eta_\mathrm{CZ} \left[1+\tanh\left(\frac{r-0.7\Rs}{0.015\Rs}\right)\right],
%\label{eq:etaE}
%\end{equation}
%where $\eta_\mathrm{CZ} = 10^{12}$~\cmss\ and $\eta_\mathrm{RZ}=0.1\eta_\mathrm{CZ}$.
%Note that this type of $\alpha$ and $\eta$ profiles are obtained from \citet{KKT06}.
With these new $\alpha$, we, however, do not get polarity reversal in a wide range of $\alpha_{0}$ 
both in solar and anti-solar DR cases.  Interestingly, this new $\alpha$ profile gives
polarity reversals if we increase the diffusivity in the bulk of the CZ. That
is, when we take $\etaSCZ = 1\times10^{12}$~\cmss, $\etasurf = 1.05\times10^{12}$~\cmss (see \Fig{etaprof} dashed line) 
and $\alpha_{0}=80$~\mps, we obtain polarity reversals in all runs except Run~E1; 
see \Fig{fig:bflyKapyla}.
It could be that in this new $\alpha$ profile, unless we increase the diffusion in the deeper CZ, 
the toroidal field generation due to shear overpowers the same due to $\alpha$ effect.
This higher diffusion makes the equatorward migration of toroidal field less important 
and the magnetic field in run E50 quadrupolar.

We note that with these new $\alpha$ profiles, we again obtain a clear increase of magnetic field
in both the sets D and E; see \Tab{table1}, although the increase is relatively small.
We note that in both sets, $\alpha$ and $\Omega$ are decreased  
with the increase of rotation period as before following \Eqs{eq:alphascale}{eq:Omegascale}.
%In Set~E, we obtain polarity reversal only in the anti-solar DR regime and the magnetic field
%does not always remain dipolar; \Fig{fig:bflyKapyla}.**** 
%We further note that in Set~E the poloidal field dominates for all the stars. This is because the $\alpha_{\phi\phi}$ is now operating in the whole CZ and thus it is more effective in generating poloidal field.

\section{Conclusion}
\label{sec:con}
In this study, using a mean-field kinematic dynamo model we have explored the features of large-scale magnetic fields of solar-like
stars rotating at different rotation periods and having the same internal structures as that
of the Sun. Our main motivation is to understand
the large-scale magnetic field generation in the slowly rotating stars with rotation period
larger than the solar value. This region is of particular interest because these stars possibly
possess anti-solar DR and thus the operation of dynamo can be fundamentally different than stars
having the usual solar-like DR.

By carrying out simulations for different stars at
different model parameters, we
show that the fundamental features of stellar magnetic field change with rotation period.
In the solar-like DR branch, we find the magnetic field strength increases with the decrease 
of rotation period. This result is, in general, agreement
with the observational findings \citep{Noyes84a,Petit08,wright11,WD16}, the global MHD convection simulations \citep{viv18,W18}, and mean-field dynamo modellings
\citep{JBB10,KKC14,KO15,Hazra19}. However, our model does not produce the observed saturation of 
magnetic field in the very rapidly rotating stars, which in the kinematic models, 
requires some additional dynamo saturation \citep{KKC14,KO15}.
In the slowly rotating stars,
we see a sudden jump of magnetic field strength at the point where the DR profile changes
to anti-solar from solar. This result is in agreement with the stellar observations
\citep{Giamp06,Giamp17,BG18} and also with the MHD convection simulations of \citet{Kar15} and \citet{W18}.
The abrupt increase of magnetic field is due to the anti-solar DR which amplifies the
existing toroidal field by supplying the same polarity field. The idea of the enhancement of magnetic field due to the change of DR was already proposed by \cite{BG18}, however,
no detailed dynamo modelling was performed.

We further show that with particular $\alpha$ profiles, the polarity reversal of the large-scale magnetic field is possible even in
the slowly rotating stars with anti-solar DR rotation provided, (i) there is a sufficiently strong
$\alpha$ for the generation of toroidal field and (ii) the anti-solar DR is non-linearly modulated
with the magnetic field such that when the toroidal field becomes strong, it quenches the shear.
Our conclusion of polarity reversal in general supports the work of \citet{viv19}, who
showed that the polarity in their global MHD convection with anti-solar DR is possible as 
the magnetic field generation through $\alpha$ effect is comparable to that of $\Omega$ effect.

One may argue that the global MHD convection simulations of stellar CZs
are still far from the real stars and there is always a question to what extent the results
from these simulations hold to the real stars. Interestingly, one robust result of these
simulations is that they all produce anti-solar DR in the slowly rotating stars with rotation
period somewhere above the solar value with Rossby number around one. Available techniques are still insufficient to confirm the existence of anti-solar DR in solar-like dwarfs; see \citet{Rei15}.
However, this has
been confirmed in some K-giants \citep{SKB03,Web05,Kho17} and subgiants \citep{Har16}. The enhancement of magnetic activity in the slowly rotating stars and its generation through the anti-solar DR give another support for the existence of anti-solar DR in slowly rotating dwarfs. Furthermore, this study along with previous observational results and global simulations suggest that the slowly rotating stars possess strong large-scale magnetic fields and possibly polarity reversals and cycles. These slowly rotating solar-like stars may also be prone to produce superflares \citep{Mae12}, which was also suggested by \citet{Kat18}.

\section*{Acknowledgements}
We thank the anonymous referee for carefully checking the manuscript and raising interesting
questions which particularly helped us to correct an error that we made in the earlier draft.
We further thank Gopal Hazra and Sudip Mandal for discussion on various aspects of the stellar dynamo.
We sincerely acknowledge financial support from Department of Science and Technology
(SERB/DST), India through the Ramanujan Fellowship awarded to B.B.K. (project no SB/S2/RJN-017/2018).
BBK appreciates gracious hospitality at Indian Institute of Astrophysics, Bangalore during the last phase of this project.
V.V. acknowledges financial support from DST through INSPIRE Fellowship.
%%%%%%%%%%%%%%%%%%%%%%%%%%%%%%%%%%%%%%%%%%%%%%%%%%



\bibliographystyle{mnras}
\bibliography{paper} % if your bibtex file is called example.bib

%%%%%%%%%%%%%%%%%%%%%%%%%%%%%%%%%%%%%%%%%%%%%%%%%%

%%%%%%%%%%%%%%%%% APPENDICES %%%%%%%%%%%%%%%%%%%%%

\appendix

\section{Supplementary material}
\label{sec:appen}
The meridional flow is obtained from the following analytical form.
%\begin{equation}
\begin{eqnarray}
v_r(r,\theta)= v_0 \left(\frac{\Rs}{r}\right)^2 \left[ \frac{-1}{m+1}+\frac{c_1}{2m+1}\xi^m- \frac{c_2}{2m+p+1}\xi^{m+p} \right] \nonumber \\
 \xi\left[2\cos^2\theta-\sin^2\theta\right]\label{mc1}
\end{eqnarray}
%\end{equation}

\begin{equation}
v_\theta(r,\theta) = v_0 \left(\frac{\Rs}{r}\right)^3 \left[-1+c_1\xi^m-c_2\xi^{m+p}\right] \sin\theta\cos\theta,\label{mc2}
\end{equation}
with
\begin{eqnarray}
 \xi(r)=\frac{\Rs}{r}-1, ~ c_1=\frac{(2m+1)(m+p)}{(m+1)p}\xi^{-m}_p, \nonumber \\
c_2=\frac{(2m+p+1)m}{(m+1)p}\xi^{-(m+p)}_p, ~ \mathrm{and} ~ \xi_p=\frac{\Rs}{r_p}-1.
\end{eqnarray}
Here $m=0.5$, $p = 0.25$, $v_0 = 10$~m~s$^{-1}$, and $r_p = 0.62\Rs$.
The boundary conditions are exactly taken from \citet{CNC04} and thus we do not repeat those here.
For the initial magnetic field, we take 
%\begin{equation}
%A=0~~~{\textrm{and}}~~~
%B=B_0 \sin(2\theta)\sin[\pi\{(r-0.55\Rs)/(\Rs-0.55\Rs)\}]
%\end{equation}
\begin{eqnarray}
A=0~~~{\textrm{and}} ~~~~~~~~~~~~~~~~~~~~~~~~~~~~\nonumber \\
B=B_0 \sin(2\theta)\sin\left[\frac{\pi(r-0.55\Rs)}{(\Rs-0.55\Rs)}\right].
\end{eqnarray}

\begin{figure}
\centering
\includegraphics[width=1.0\columnwidth]{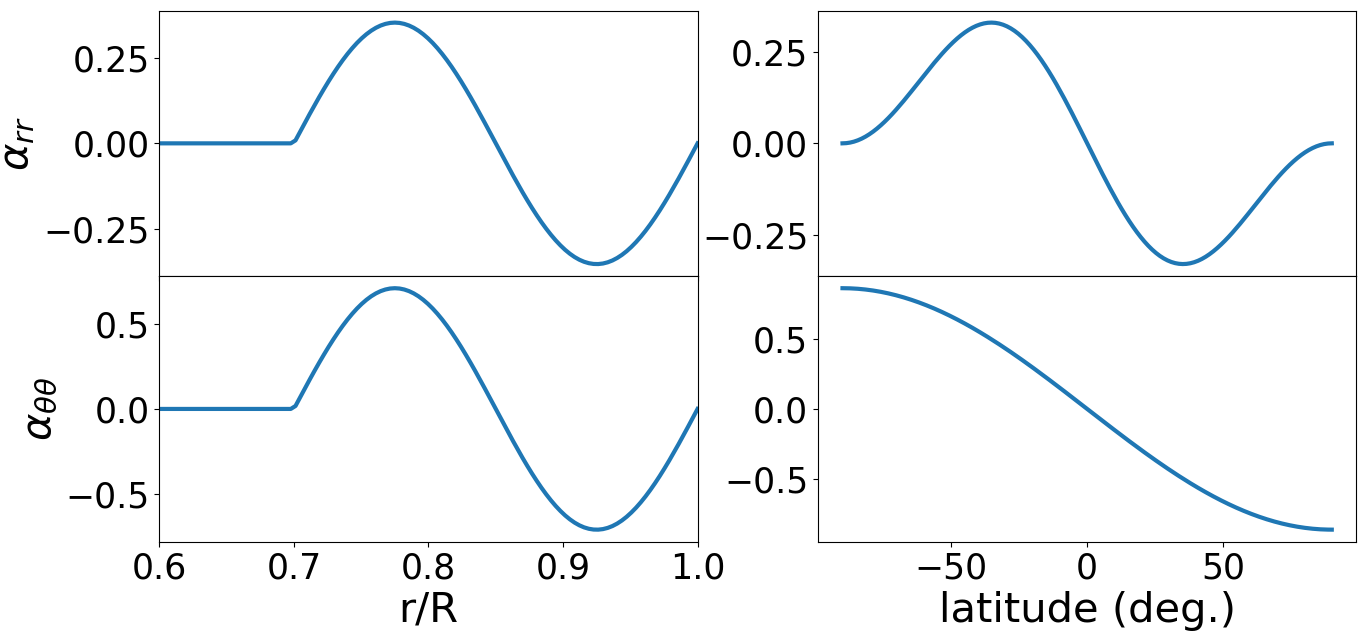}
\caption{
Left: Radial variations of $\alpha_{rr}$ and $\alpha_{\theta\theta}$ in \mps\ at $45^\circ$ latitude.
Right: Latitudinal variations of the same at $r=0.95\Rs$. 
These are the profiles used in Set D; \Eq{eq:altD} with $\alpha_0 = 1$~\mps.
}
\label{fig:alphaD}
\end{figure}

\begin{figure}
\centering
\includegraphics[width=.7\columnwidth]{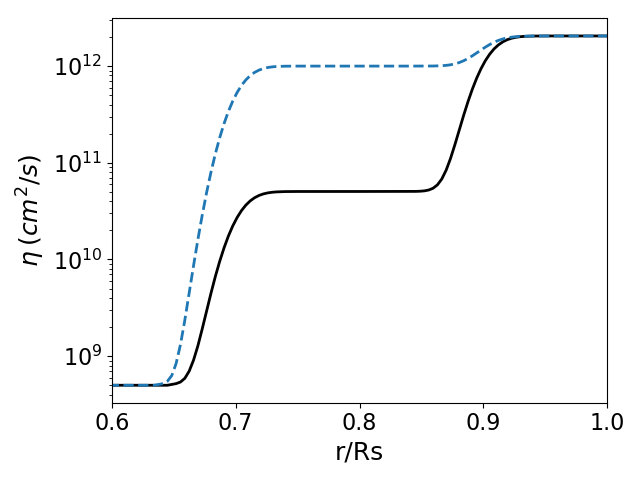}
\caption{
Solid line: diffusivity profile $\eta$ used in all the sets of simulations, except Set E for which the 
profile shown by the dashed line is used.
}
\label{etaprof}
\end{figure}

\begin{figure}
\centering
\includegraphics[width=1.\columnwidth]{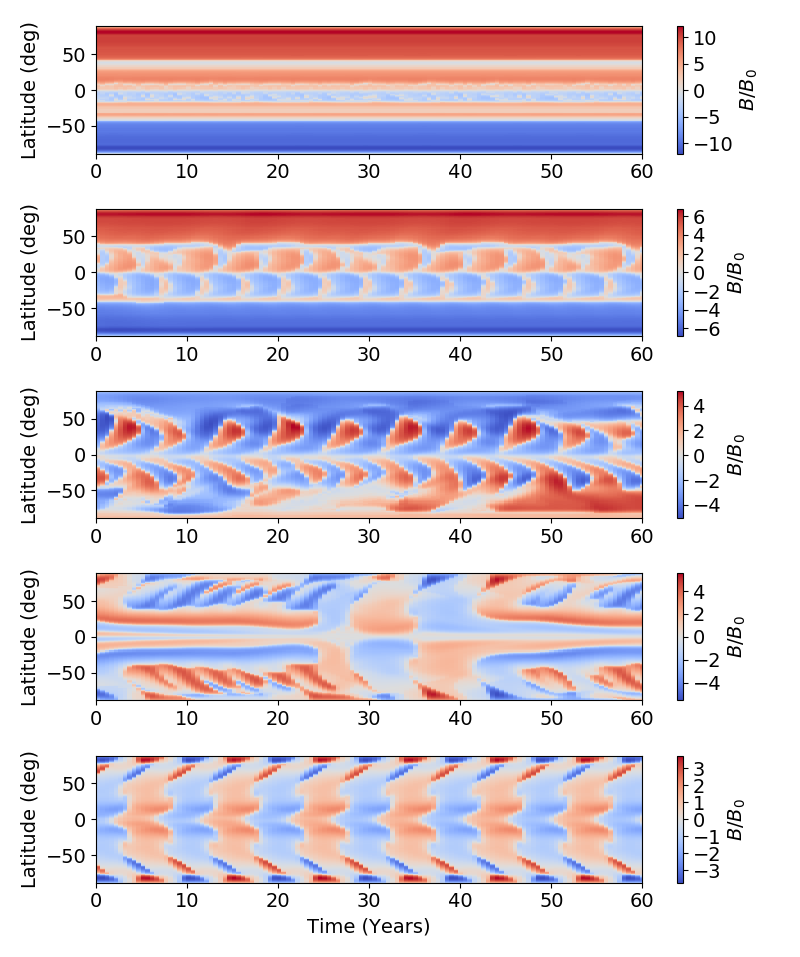}
\caption{Same as \Fig{fig:allbfly}, but obtained from Runs~E5, E17, E25, E32, E50 (top to bottom).
}
\label{fig:bflyKapyla}
\end{figure}

%%%%%%%%%%%%%%%%%%%%%%%%%%%%%%%%%%%%%%%%%%%%%%%%%%

% Don't change these lines
\bsp	% typesetting comment
\label{lastpage}
\end{document}